# Synergistic Interface Effects in Composite Dielectrics: Insights into Charge Trapping Regulation through Multiscale Modeling

Haoxiang Zhao[1,2,3], Lixuan An[4], Daning Zhang[1], Xiong Yang[1], Huanmin Yao[1], Guanjun Zhang[1], Haibao Mu[1]* and Björn Baumeier[2,5]*

1 State Key Laboratory of Electrical Insulation and Power Equipment, School of Electrical Engineering, Xi'an Jiaotong University, Xi'an 710049, China

2 Department of Mathematics and Computer Science, Eindhoven University of Technology, Eindhoven 5600MB, The Netherlands

3 Department of Applied Physics, Eindhoven University of Technology, Eindhoven 5600MB, The Netherlands

4 KERMIT, Department of Data Analysis and Mathematical Modelling, Ghent University, Ghent 9000, Belgium

5 Institute for Complex Molecular Systems, Eindhoven University of Technology, Eindhoven 5600MB, The Netherlands

*Corresponding author

E-mail address:

haibaomu@mail.xjtu.edu.cn (Haibao Mu)

b.baumeier@tue.nl (Björn Baumeier)

**Abstract**

The rapid development of modern energy applications drives an urgent need to enhance the dielectric strength of energy storage dielectrics for higher power density. Interface design is a promising strategy to regulate the crucial charge transport process determining dielectric strength. However, the targeted exploitation of interface effects on charge transport is limited due to a lack of fundamental understanding of the underlying mechanisms involving elementary electronic processes and details of the intricate interplay of characteristics of molecular building blocks and the interfacial morphology – details that cannot fully be resolved with experimental methods. Here we employ a multiscale modeling approach linking the quantum properties of the charge carriers with nano- and mesoscale structural details of complex interfaces. Applied to a prototypical application-proven cellulose-oil composite with interfaces formed between oil, disordered, and crystalline cellulose regions, this approach demonstrates that charges are trapped in the disordered region. Specifically, it unveils this trapping as a synergistic effect of two transport-regulating interface mechanisms: back-transfer to the oil region is suppressed by energetic factors, while forward-transfer to the crystalline cellulose is suppressed by low electronic coupling. The insight into the molecular origins of interface effects via dual-interface regulation offers new development paths for advanced energy materials.

# 1. Introduction

Driven by the rapid development of renewable power, electrified transportation, and electroactive actuators, the dramatically increasing demand for energy storage dielectrics with high power density urgently necessitates significant improvements in dielectric strength[1-7]. Since T. J. Lewis's milestone introduction of the "nanodielectrics" concept in 1994, interface effects have been recognized as crucial in improving dielectric strength[8-11], and interface design has been regarded as an effective practical strategy[12-15]. Although it is widely acknowledged that the dielectric strength is primarily influenced by charge transport processes[8-11,16-18], many of its improvements are attributed to the hindrance of charge transport[19-38], the field lacks a fundamental understanding of interface effects on charge transport, its underlying mechanisms, and even the basic microscopic dynamic properties. Consequently, practical design rules for guiding material design are speculative at best.

Nevertheless, progress has been made in addressing practical issues in developing and engineering of interfaces in polymer-based energy storage dielectrics design[12-15]. This includes various nano-doping techniques[12,19-23] and hierarchical interface designs[12,24-28]. However, these applications largely remain trial-and-error[13,33], and a unified strategy has yet to emerge. Stronger yet, inconsistent or contradictory explanations of the interface effect mechanism are also reported[13]. Mainstream theoretical models based on the electrical double-layer model suggest that accumulated charge inhibits subsequent charge transport[8-10,39]. However, its premise of relatively strong charge conduction is questionable in polymer composite dielectrics, where charge transport typically occurs via hopping across energy barriers associated with localized electronic states due to the lack of long-range crystalline order[40-42]. Particularly, Heeger presented in his Nobel lecture the understanding of the hopping-type charge transport in polymeric and organic materials referring to the Marcus theory[43,44], which is expected to be a promising theoretical framework for the charge transport in polymer dielectrics. Yet the associated popular arguments involving the energy bandgap and electrostatic potential are not directly related to hopping-type charge transport in localized-state polymer dielectrics[40-42]. Bandgap considerations pertain to semiconductors and crystalline materials where the charge carriers move freely after being excited across the gap. Arguments based purely on the electrostatic potential relate to (macroscopic) charge distributions in a static picture.

Fundamentally, challenges in understanding interface effects on charge transport stem from the limitations in resolving the dynamic charge transport process from the nano- to macroscale (or at least mesoscale) and thereby in shedding light on the interplay between the chemistry of the molecular building blocks and the local and global structural features (morphology) of the interfaces. On the one hand, state-of-the-art experimental techniques, like pulsed electric-acoustic current or thermal simulated depolarization current, can only reflect the whole sample's overall charge distribution or trap characteristics without microscopic dynamic details[13,18,37,38,45,46]. Based on these macroscopically measured bulk properties of the material, it is difficult to even distinguish between the contributions from the newly added component itself or the created interface. On the other hand, quantum chemistry (QC) calculations, e.g., based on Density Functional Theory (DFT), offer nanoscale electronic structure insights[6,13,35,37-39,45,46] but fail to capture charge dynamics in a complex interface structure at a larger scale. The limited understanding of charge transport characteristics results in interface effects being treated as a black box problem, leading to speculative or hypothetical theoretical models at best. A bridge connecting dynamic charge transport and structural material properties across scales is urgently needed[13,47].

Multiscale modeling offers such a bridge to provide unprecedented insight into the interface effects on charge transport[48-50]. Combining molecular dynamics (MD) simulations to obtain atomic-level resolution of an interface morphology and large-scale embedded quantum chemistry calculations, our bottom-up multiscale approach allows for a predictive evaluation of the electronic properties of the individual molecular building blocks and their interactions as they enter the Marcus rate for electron-transfer theory. With this, parameter-free charge transport simulations are realized within a realistic morphology as rate-based dynamics using kinetic Monte Carlo (KMC) methods. Such a combined approach covers the interplay of the two key aspects – realistic interface structure and dynamic charge transport – promising an understanding of interface effects from nano to macro/mesoscale.

To demonstrate the advantages of this approach, we are using the classic cellulose-oil composite as a study subject. Its dielectric strength is significantly increased by the cellulose-oil interface introduced through the simple impregnation of porous cellulose dielectric paper with oil[32,51,52]. Widely used in electrical engineering, its reliability stands the test of several decades during the electrification process[7,53]. Through multiscale modeling, this work aims to grasp the characteristics of interface-related charge transport, explore the mechanisms of interface effects on charge transport regulation, and offer practical guidance for dielectrics interface design. To this end, firstly, a representative cellulose-oil interfacial morphology is modeled on atomic resolution, containing two typical interface types in dielectrics design: one between oil and disordered hemicellulose and one between hemicellulose and crystalline cellulose. Then, a combination of quantum and classical methods (see "Methods" section ) is used to evaluate the physical quantities entering the Marcus hopping rate, i.e., the ionization energy of each molecular unit (measuring its charge-attracting ability), also known as *site energy*, and the strength of the *electronic coupling* between neighboring units, as they are influenced by structural disorder. Finally, the charge dynamics are obtained on the whole oil-cellulose system and further analyzed in terms of the influence of site energy differences, external driving force, and electronic coupling for both interface types.

Results unveil remarkable trap characteristics of the cellulose-oil interface: once the charge enters the region between two interfaces, it hardly escapes by either back-transfer recrossing the interface 1 or forward-transfer crossing the interface 2 to the crystalline cellulose. The trapping of charges in the hemicellulose region is, therefore, efficiently regulated by the synergy of both interfaces rather than by a single interface. Detailed analysis also reveals different regulatory mechanisms at the two interfaces, where the trapping at the oil-hemicellulose interface is regulated by differences in site energies, whereas the trapping at the hemicellulose-cellulose interface is regulated by low electronic coupling across it. While energy-regulated trapping is a widely recognized concept, its synergy with the crucial role of coupling-regulated trapping in the emergence of what is referred to as the interface effect has previously been overlooked.

Unveiling and then utilizing the synergy between these two regulating mechanisms and their molecular origin allows for a more flexible approach to interface engineering. The insights gained from our findings not only broaden the view on the origins of the interface effects, thereby offering more reasonable explanations for various successful advanced material modifications[33-38], but may unify related dielectric material innovation at a higher level.

## 2. Results

### 2.1. Cellulose-oil Interface and Multiscale Model

The first step in the multiscale model of charge dynamics for localized-state polymer dielectrics is the simulation of a realistic, representative morphology of a cellulose-oil interface with MD. The structural model we adopt here is based on multiscale deconstruction analysis[54-57] as shown in Fig. 1a: From the cellulose dielectric paper, macrofibrils can be observed at the 100 μm scale. Each macrofibril is composed of repeating microfibrils of 1 μm scale. Ultimately, microfibrils consist of cellulose molecular chains at the nanoscale. In detail, the core of the microfibril is crystalline cellulose, surrounded by disordered hemicellulose (Fig. 1b). The molecule structures of cellulose and hemicellulose are shown in Fig. S1. The final constructed cellulose-oil interfacial model includes oil and repeating microfibrils (left part in Fig. 1c), and computational details on MD modeling are provided in the "Methods" section. There are two typical interfaces along the electric field. Interface 1 is between oil and disordered hemicellulose, which can represent an interface formed by two different materials and the mechanism subsequently revealed can apply to the nano-doping interface. Interface 2 is between disordered hemicellulose and crystalline cellulose, which can represent an interface formed by two similar materials and the mechanism can apply to the interface in all-organic composites.

After the interfacial morphology model is simulated, it is partitioned into hopping sites (segments/monomers) for charge transport simulation. The charge dynamics model starts from a single charge hopping (or electron transfer event) between a pair of sites integrating factors of different scales, and builds charge transport in the interfacial morphology as a sequence of such bi-molecular transfers. The logic flow is as shown from ① to ④ in the right part of Fig. 1c. Based on the Marcus theory for hopping-type charge transfer in localized-state dielectrics[40-44], the bi-molecular hopping rate $\omega_{ij}$ between two sites $i$ and $j$ can be calculated according to

$$\omega_{ij} = \frac{2\pi}{\hbar} \frac{J_{ij}^2}{\sqrt{4\pi\lambda_{ij}k_{\mathrm{B}}T}} \exp\left[-\frac{(\Delta E_{ij} - \lambda_{ij})^2}{(4\lambda_{ij}k_{\mathrm{B}}T)}\right] \quad (1)$$

and depends on two crucial factors: the site energy difference $\Delta E_{ij}$ and the electronic coupling $J_{ij}$ (other parameters in equation (1) are introduced in the "Methods" section). The former is determined by the individual charge-attracting abilities (ionization potentials) or site energies of the monomers $E_i$, which comprise internal quantum-mechanical (from the chemistry of the material) and external electrostatic contributions (from the morphology), as well as contributions from an external electric field. The detailed calculation process will be introduced in section 2.3. The electronic coupling $J_{ij}$ 'bridges' two monomers and is a measure of the quantum-mechanical interaction between site-localized electronic states. It will be analyzed in section 2.4. Based on the Marcus theory, not only the multiscale factors can be integrated during the calculation of hopping rates, but also the charge transport in localized-state dielectrics is discussed in a complete and appropriate theoretical framework.

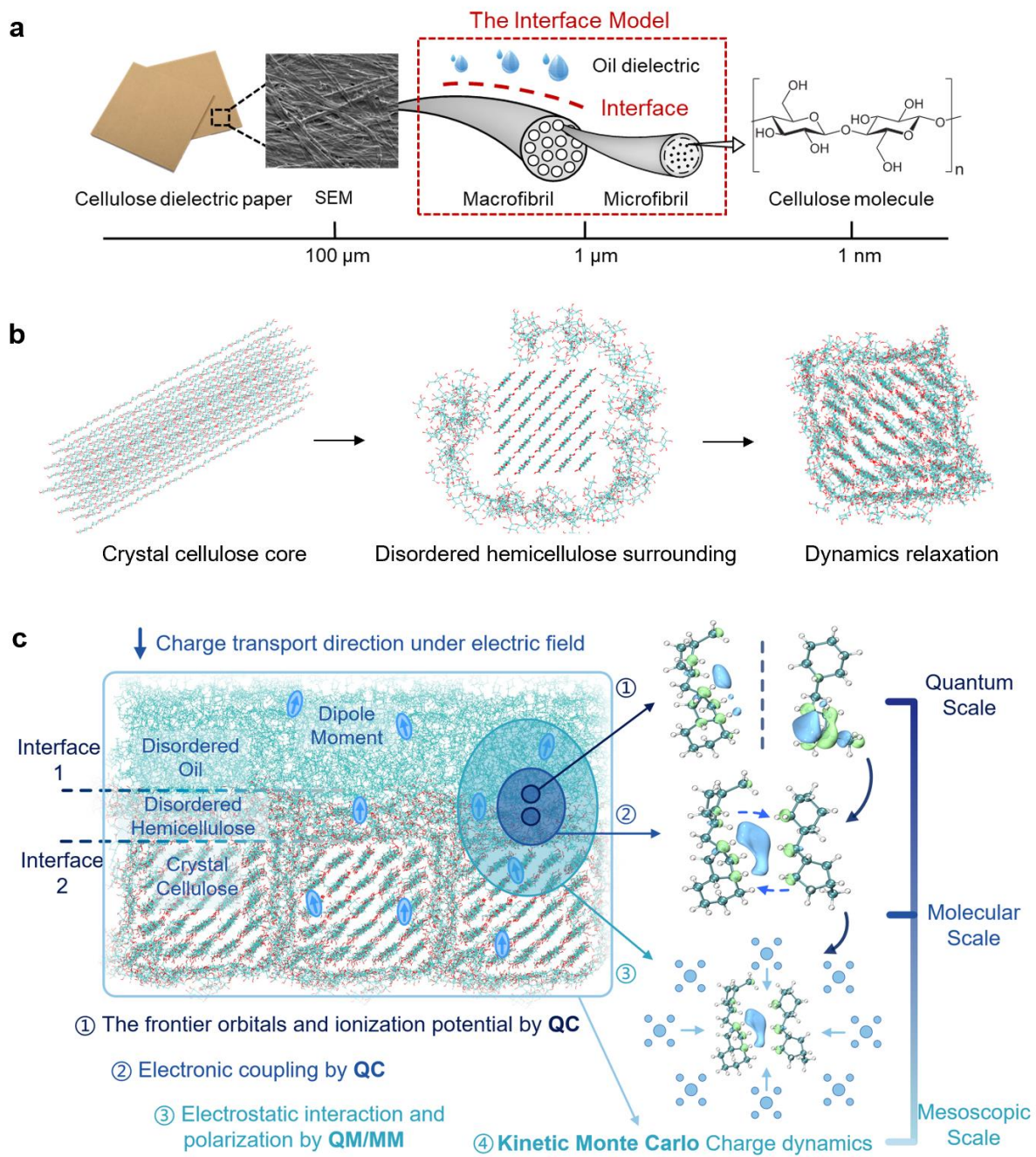

Figure 1. Cellulose-oil interface multiscale modeling. **a** Conceptualization of the cellulose-oil interfacial model. **b** MD model of cellulose microfibril, with a crystalline cellulose core surrounded by disordered hemicellulose. **c** Schematic of the elements of the multiscale method in the constructed interfacial morphology.

## 2.2. Charge trapping and its regulation by a dual-interface effect

With all the charge hopping rates, charge hopping trajectories in the morphology are obtained. Fig. 2a shows three detailed examples of individual KMC charge transport trajectories based on 100 hopping steps, where each polyline segment represents the hopping trajectory between two specific hopping sites. Fig. 2b presents a heat map of normalized count statistics of involved sites based on 5000 KMC trajectories. Injected into the oil region, the charges hop along the electric field with a spreading trend due to the diffusive, random walk nature of the hopping-type charge transport. The charges easily hop across interface 1 into the middle disordered hemicellulose region but are effectively trapped there: escaping back across interface 1 to the oil region or forward across interface 2 into the crystalline cellulose is blocked. Trapped charges transfer between sites within the middle region between two interfaces, leading to a deeper trajectory color in the heat map.

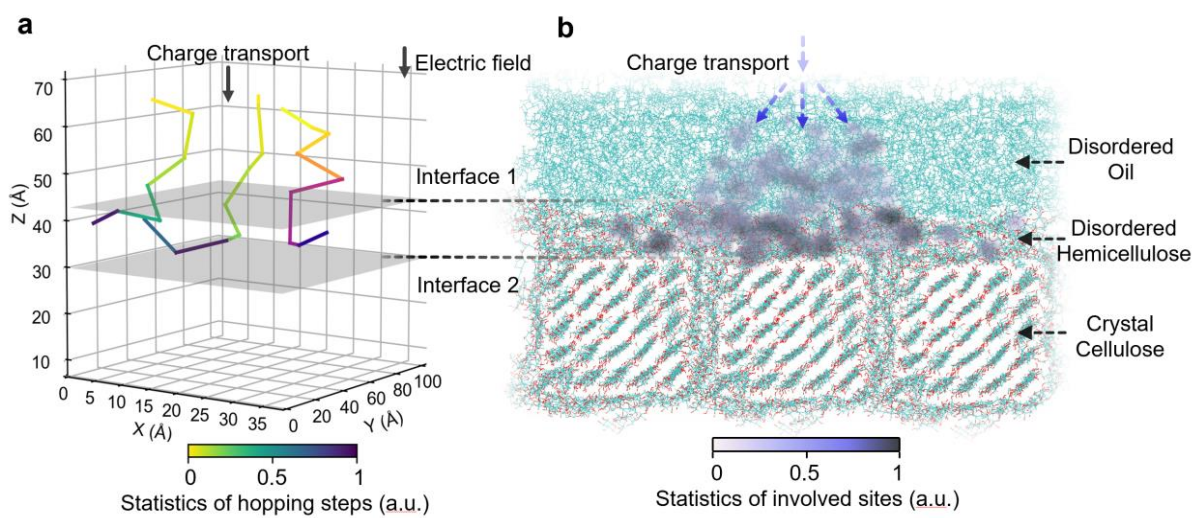

Figure 2. Intuitive charge transport dynamics based on kinetic Monte Carlo to show charge transport regulation by dual-interface effect. **a** Three detailed examples of charge transport trajectories. **b** Heat map of 5000 charge transport trajectories. The electric field is 50kV/mm directing from the oil region to the cellulose region.

The charge trajectories from our multiscale approach explicitly reveal for the first time a dual-interface effect regulating charge trapping microscopically. This dual-interface effect can also be seen in more detail by an inspection of the interface-related characteristics of the charge hopping rates $\omega_{ij}$. In Fig. 3, we show rate distributions classified according to the involved regions: oil, hemicellulose, and crystalline cellulose, respectively, including directionality across interfaces, as indicated in the schematic diagram at the top of the figure. First, Fig. 3a shows the hopping rates when the charge is in front of interface 1 and we focus primarily on the high-rate subset of the distributions. The hopping rates of the Oil→Hemi processes are overall similar to the Oil→Oil related ones, with a few slightly higher values. This indicates that the charge transport inside the oil and from the oil into the hemicellulose is comparable and that there is effectively no local barrier for crossing the interface in the direction of the eternal field. When the charge is in the hemicellulose region between two interfaces, the hopping rates of Hemi→Hemi are several orders of magnitude larger than Hemi→Oil, as shown in the red dotted box in Fig. 3b. Therefore, after the initial injection of the charge into the hemicellulose, transport within this region is preferred and the transfer back to the oil region unlikely.

Focusing now on the processes involving interface 2, one can identify the opposite effect compared to the situation at interface 1. As one can see from the high-rate data in Fig. 3c, hopping rates for crossing from the hemicellulose into the crystalline region are significantly smaller than for transfer processes within hemicellulose. Consequently, not only is the back-transfer of the charges from the hemicellulose to the oil region across interface 1 suppressed, but also the forward-transfer along the direction of the electric field across interface 2.

In summary, the result of the interplay of both interfaces is an effective trapping of the charges in the hemicellulose region.

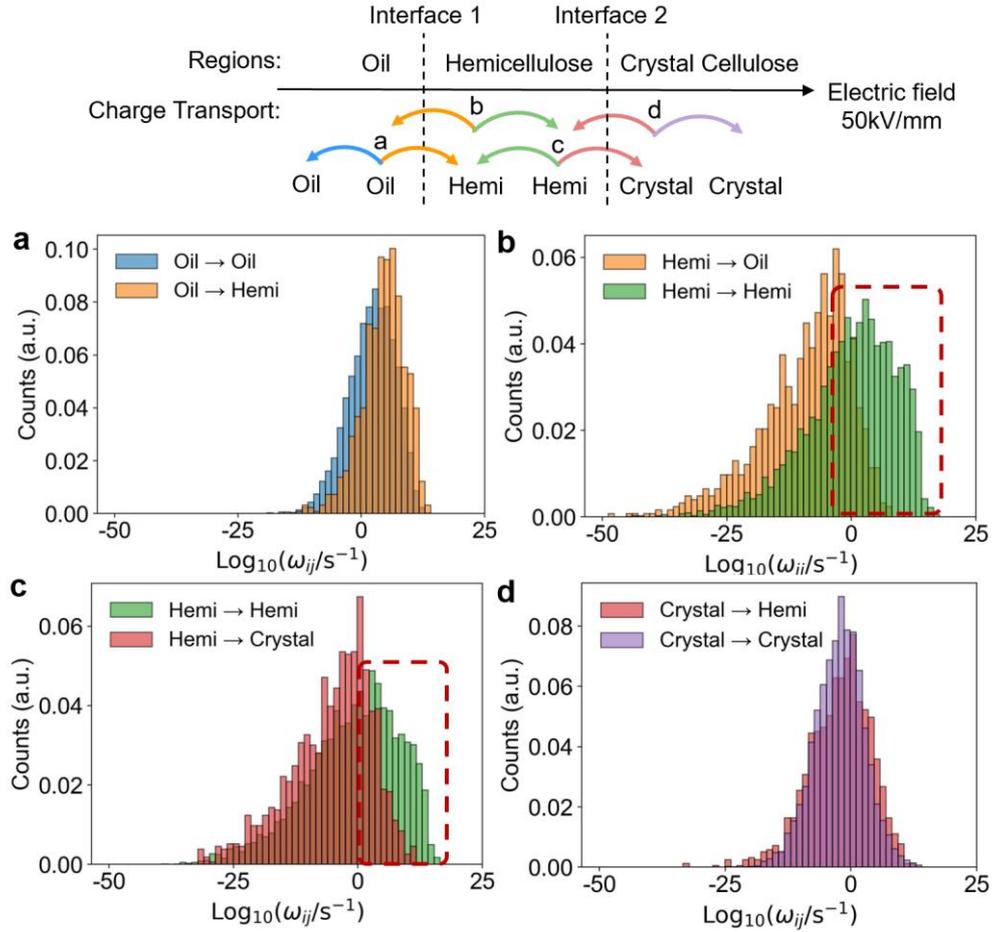

Figure 3. Statistical region-based results of charge hopping rate $\omega_{ij}$. **a**, **b** Results related to interface 1. **c**, **d** Results related to interface 2. The electric field is 50kV/mm directing from oil region to cellulose region.

More than a simple blocking of charge transport by a single interface, this is a dual-interface charge transport regulation realized by the synergy of the two interface effects. This regulation of charge trapping is crucial for dielectric materials and aligns with the long-standing goals of dielectrics design. Compared to the trapping site introduced by previous molecular unit or group modification methods, this tailored region between two interfaces can provide more charge trapping sites, thus realizing a more effective charge transport regulation. Practically, this dual-interface effect matches the two mainstream interface implementation methods and can be correspondingly constructed by surface modification of nanoparticles[12,19-23] and tailoring multi-layer structures[12,24-28]. Representing two generalized interface types presenting broad application potential, the revealed two interface effects are worth further studies of the underlying mechanisms. The following in-depth exploration of these two interface effects is based on two key aspects of charge hopping rate: charge attracting ability measured by the ionization energy of individual molecules and the electronic coupling between two.

**2.3. Characterization of charge attracting ability and driving force for charge hopping**

The charge attracting ability or ionization energy of a segment/molecule is discussed often discussed in terms of frontier orbital energies obtained from effective single-particle quantum chemistry methods such as DFT[6,13,35,37-39,45,46]. However, it is well known that even for single molecules, these estimates are quantitatively and sometimes also qualitatively inaccurate. Single molecule calculations also neglect the modification of the ionization energies due to the presence of other molecules in a morphology such as the oil-cellulose interface. Frontier orbital energies are also by definition proxies for vertical ionization energies, while the Marcus rate as in equation (1) requires the difference of adiabatic energies in $\Delta E_{ij}$. Instead, we here evaluate taking hole transport as an example the ionization potential, i.e., the energy required to remove an electron from a neutral molecule[41,43] from the difference of total energies of the molecules in charged ($E_{charged}$) and neutral ($E_{neutral}$) states:

$$\begin{aligned} E = IP &= E_{charged} - E_{neutral} \\ &= IP^{int} + IP^{env} \end{aligned} \quad (2)$$

In equation (2), we have additionally split the ionization potential into internal (single molecule) and environment contributions. The schematic is shown in Fig. 4a. In the previous reported studies[6,13,35,37-39,45,46], the charge attracting ability is characterized only by the original frontier orbital energy calculated by DFT, in which cases $-\varepsilon_{HOMO}$ is applied as $IP^{int}$. Here, we first get a proper $IP^{int}$ from the calculations of DFT total energies in both states. Then, multiple environment factors are accounted for in $IP^{env}$ by a classical, atomistic model including electrostatic interactions and polarization effects. Taken together, the site energy is therefore calculated in a quantum-classical (quantum mechanics/molecular mechanics, QM/MM) setup whose details can be found in "Methods" section 4.3.

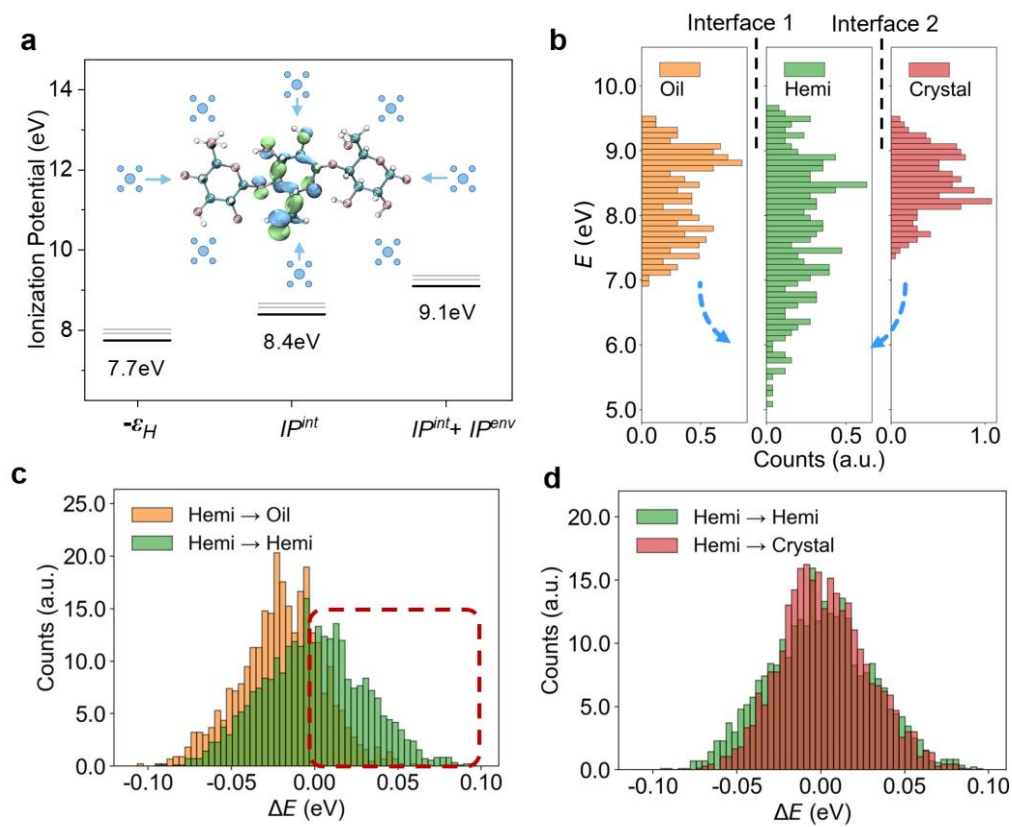

Figure 4. Characterization of charge attracting ability by site energy and interfacial analysis. **a** Schematic of the QM/MM calculation process of site energy. **b** Statistics of site energy $E$ of different regions. **c** Region-based statistics of site energy difference $\Delta E$ related to interface 1. **d** Region-based statistics of site energy difference $\Delta E$ related to interface 2. No external electric field contribution is involved in the results of **a**, **b**, **c**, **d**.

The distribution of site energies, or the density-of-states (DOS) for charge transport, calculated according to equation (2) are shown split into the three regions of the oil-cellulose interface in Fig. 4b. To show the properties of the material itself, the results do not include an externally applied electric field, which will be studied later. The hole charge carriers energetically favor relaxation to low values of $IP$. In general, the region containing lowest site energy is in hemicellulose region, and from purely energetic considerations on the respective DOS this seems to indicate that charge carriers can easily hop into but difficult to escape from hemicellulose region. This suggests that the trapping of charge carriers in the hemicellulose region is purely energy-regulated across both interfaces.

However, inspecting the overall DOS is misleading in the context of hopping transport as the fundamental electron transfer process is short-ranged and driven by the energy difference among hopping pairs. It is therefore more instructive to consider the distribution of site energy differences, or driving forces, $\Delta E_{ij}$ as it enters the Marcus rate in equation (1).

Fig. 4c and Fig. 4d show the site energy difference $\Delta E$ related to interfaces 1 and interface 2, corresponding to Fig. 3b and Fig. 3c, respectively. Fig. 4c shows that in the relevant part of positive driving force, site energy difference between Hemi and Hemi is indeed larger than the one between Hemi and Oil, as shown in the red dotted box. The stronger driving force to remain in the hemicellulose region than to transfer back to oil supports with the conclusions drawn from the overall charge hopping rates shown in Fig. 3b for interface 1, and it confirms that the trapping of charges on this interface is indeed energy-regulated.

However, at interface 2 between hemicellulose and crystalline cellulose, the respective driving forces distributions of the related charge transport as shown in Fig. 4d are similar. This lack of an effective energetic barrier is surprising given the pronounced differences in the respective rates in Fig. 3c and excludes energy-regulation as a mechanism for trapping the charge carriers in the hemicellulose region at interface 2. In summary, the results in Fig. 4c and Fig. 4d indicate that the site energy difference significantly contributes to the interface effect of interface 1 but not much to interface 2. To explain the interface effect at interface 2, we will consider the often-overlooked electronic coupling in section 2.4.

Before turning to this analysis, in the context of driving forces, we briefly discuss the model of a shielding effect by charge accumulation, which is currently a popular explanation for the interface effect. When charges get trapped (or at least slowed down) at the interface, their presence creates an additional electric field that affect the other dynamic carriers. Instead of simulating these effects explicitly, we vary the interface-associated electric field during site energy difference calculations (the related principle can be found in "Methods" section). Fig. 5a shows a comparison between the total driving force in the whole system for an external field of 50 kV/mm and the contribution of the external field, demonstrating that the electric field is not the dominant factor of in the site energy difference. In Fig. 5b and Fig. 5c we focus on interface 1 and interface 2 respectively and strengthen (500 kV/mm) or weaken (5kV/mm) electric field during the calculation of interface crossing rates. With the increasing of the electric field, the rate distributions shift towards lower values when the hopping is back-crossing in Fig. 5b, and shift towards higher values when the hopping is forward-crossing in Fig. 5c. It should be noted that the electric field effect is significant itself as the rates shifting is shown in the logarithmic coordinates. But the relative distribution differences between Hemi-Hemi and Hemi-Oil or between Hemi-Hemi and Hemi-Crystal are not dramatically affected by the electric field. This implies that the shielding effect is insufficient to explain the interface effect, at least the observed trapping behavior in our interface system, for which the energetic factor discussed above could be more important.

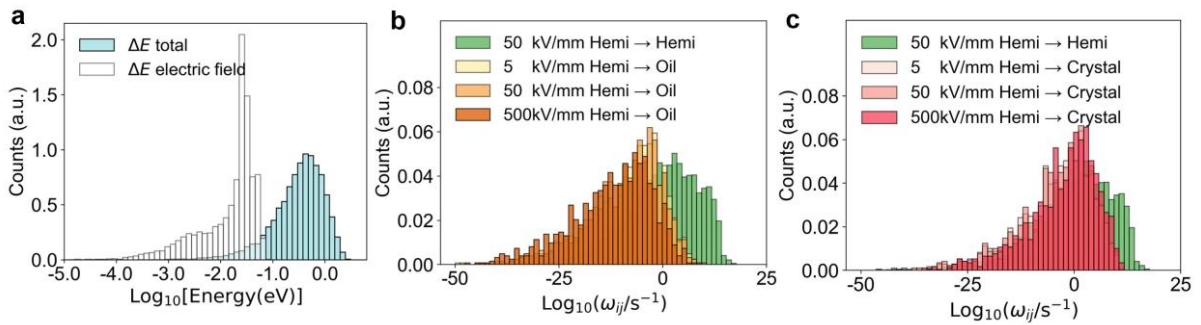

Figure 5. Influence of electric field on charge hopping rate related to interface. **a** Comparison of the electric field component of ΔE and the entire ΔE. **b** Influence of electric field on hopping rate related to interface 1. **c** Influence of electric field on hopping rate related to interface 2.

## 2.4. The crucial role of molecular electronic coupling for interface effects

Beyond the driving force of site energy difference $\Delta E$, electronic coupling $J_{ij}$ in equation (1) is the second key quantity influencing charge hopping rates. It measures the quantum-mechanical interactions of the localized electronic states identified with the frontier orbitals of two hopping sites, acting as a 'bridge' for charge hopping, as the HOMO isosurface of the dimer shown in Fig. 1c ②. Despite its essential position, electronic coupling is often overlooked in the previous related studies where a comprehensive theoretical framework for charge transport is lacking. The essence of the interface effect on charge transport is the interplay between the nanoscale electronic process and the material morphology structure, therefore the electronic coupling is the key to deconstructing this multiscale interplay.

Using an efficient method, we calculate the electronic coupling for all possible charge transport site pairs in the morphology. The basic calculation principle is described in the "Methods" section. Electronic coupling is highly sensitive to molecule types, relative orientation and distance[58,59]. Fig. 6 highlights this sensitivity with a specific group of pairs, where the centered highlighted molecule is the fixed component of all pairs, while the surrounding two layers of molecules with blue or grey color are the second components of pairs. Despite visually similar distances between the components of pairs ①, ②, and ③, differences in relative orientation and molecule types result in electronic coupling values varying by two orders of magnitude. Across the interface morphology, results span nineteen orders of magnitude, indicating the electronic coupling's significant influence on charge transport and its expected critical role in interface effects.

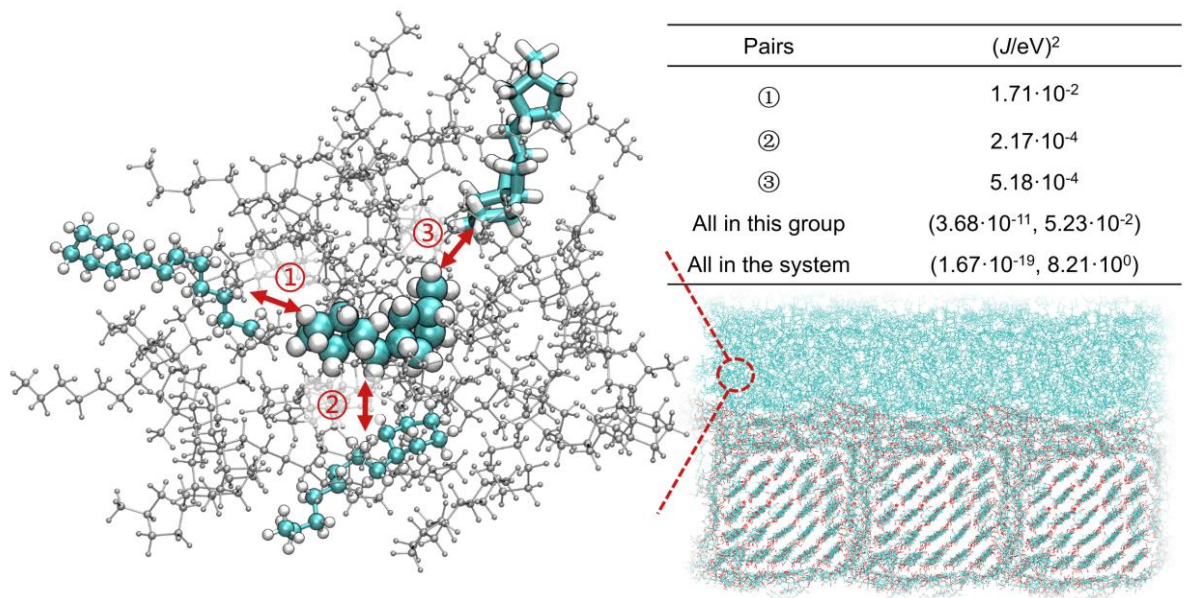

Figure 6. Sensitivity of electronic coupling. In the example group on the left side, pair ① and ② are consist of the same types of molecules but with different relative orientations, pair ① and ③ are consist of different types of molecules. The right side is the summary of electronic coupling values.

Fig. 7 presents the distributions results of electronic coupling $J$. Fig. 7a and Fig. 7b are related to charge hopping across interface 1, while Fig. 7c and Fig. 7d are related to interface 2. The region-based results align with charge hopping rate statistics in Fig. 3. Fig. 7b shows that the electronic coupling for hopping across interface 1 is significantly lower than the one for hopping within the middle disordered hemicellulose region between two interfaces. This difference, combined with the site-energy differences discussed above, leads to a hindrance of charge back-transfer across interface 1, and favors transfer inside the disordered region. Fig. 7c shows that the electronic coupling for hopping across interface 2 is also greatly lower than the one for hopping within the middle region. Here, however, this low coupling is crucial to explain the significantly lower rates for charge transfer into the cellulose crystal and the effective trapping of charges at the interface. Considering the weak contribution of site energy differences to the interface effect of interface 2 shown in Fig. 4d, it is clear that charge trapping at interface 2 is coupling-regulated.

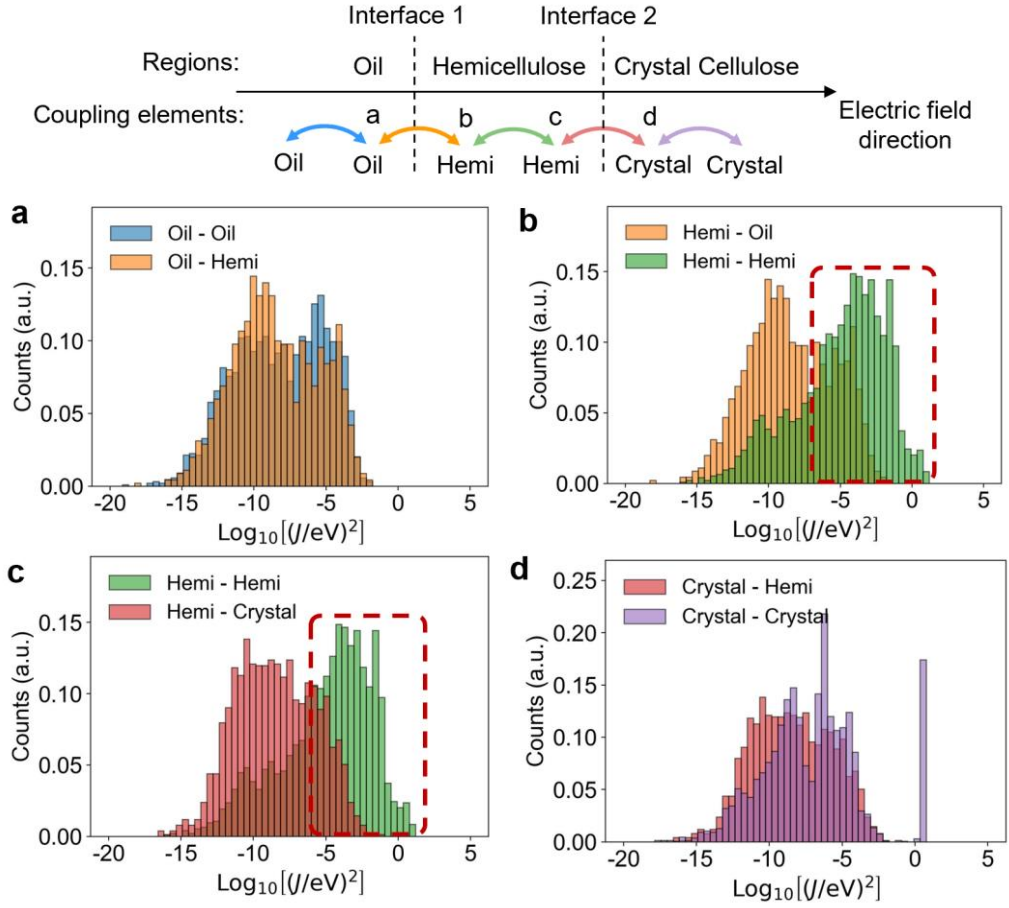

Figure 7. Region-based statistical results of electronic coupling $J$. **a**, **b** Results related to interface 1. **c**, **d** Results related to interface 2.

This notion of coupling-regulation vs energy-regulation is further corroborated by analyzing the different factors of the Marcus rate in equation (1): the exponential term $G_{ij}(\Delta E_{ij})$ and the remaining factor $F_{ij}(J_{ij})$, representing contributions of $\Delta E_{ij}$ and $J_{ij}$ to $\omega_{ij}$, respectively. The detailed definitions of $G_{ij}(\Delta E_{ij})$ and $F_{ij}(J_{ij})$ are in the Supporting Information, followed by the statistical region-based results and the correlation analysis to the hopping rates shown in Fig. S2 to Fig. S5. The general distribution characteristics of $\Delta E_{ij}$ contribution $G_{ij}(\Delta E_{ij})$ is similar to the results of single $\Delta E_{ij}$, while the correlation between this contribution and the hopping rates is strong. On the other hand, the general distribution characteristics of the coupling factor $F_{ij}(J_{ij})$ are similar to the results of single $J_{ij}$, while there is no obvious correlation between this contribution and the hopping rates. By comparing the $\omega_{ij}$ in Fig. 3 and the $J_{ij}$ contribution $F_{ij}(J_{ij})$ in Fig. S5, it is worth noting that the value range of $F_{ij}(J_{ij})$ is much narrower.

Simultaneously considering the determining effect of $\Delta E_{ij}$ on the specific distribution characteristics and its basic position of $\omega_{ij}$, it can be found that besides $\Delta E_{ij}$, $J_{ij}$ plays a role in changing the mean order of magnitude of the hopping rate, intuitively shifting the position of $\omega_{ij}$ of each individual region as a whole. For example, in Fig. 3c, the interface effect of interface 2 is directly reflected by the overall difference in relative position between the hopping rate from Hemi to Hemi (with green color) and the hopping rate from Hemi to Crystal (with red color). This relative position difference of hopping rate is primarily attributed to the shifting effect of electronic coupling (Fig. 7c) rather than site energy difference (Fig. 4d). Therefore, it can be concluded that the electronic coupling dominates the interface effect of interface 2 on charge transport, while it amplifies the interface effect of interface 1, which is dominated by site energy difference.

## 3. Discussion

The intuitive microscopic charge transport dynamics, clear hopping rates and comprehensive analysis above provides valuable insights into the interface effects on charge trapping regulation that significantly improve the dielectric strength of cellulose-oil composites. The corresponding mechanism is not merely a single interface effect that inhibits charge transport, but a dual-interface effect from the synergy of two different interface effects, resulting in a charge trapping region between two interfaces. Utilizing Marcus theory for localized-state polymer dielectrics, it is discovered that the interface effects are regulated to two crucial factors: charge attracting ability or ionization energy and molecular electronic coupling. For the oil-hemicellulose interface (interface 1), both two factors contribute to the interface effect, although the energy-regulation is the dominant mechanism here. However, for the hemicellulose-cellulose interface (interface 2), the energetic driving forces for transfer within hemicellulose and transfer across the interface to the cellulose crystal are quite similar due to the similar monosaccharide building blocks. Here, electronic coupling is the pivotal factor for the charge hopping difference between the two interface components, leading to a different, coupling-regulated interface effect. Eventually the interaction of the two interfaces leads to efficient charge trapping. The energy-regulated and coupling-regulated mechanisms of these two typical interface effects are broadly applicable to help with the explanation for two different types of interfaces, which are the interface with different components such as nano-doping interface, and the interface with similar components such as interface in all-organic composite dielectrics.

The further detailed analysis of site energy and electronic coupling offers robust theoretical support for the mechanisms of improving dielectric strength. The importance of charge attracting ability has previously shown practical effects in improving dielectric strength, exemplified by the introduction of high-electron-affinity molecular semiconductors[34] or electronegative molecular groups[38]. On the other hand, and more importantly, recognizing the crucial role of molecular electronic coupling for hopping-type charge transport in complex polymer-based dielectrics fills a significant theoretical gap. For instance, research on spiral-structured dielectric polymers exhibiting ultra-high performance explains the regulation of charge transport based on free volume[35] seems related as free volume can be associated with vanishing electronic coupling because it decays exponentially with distance. Our work comprehensively presents the concept, characteristics, and importance of electronic coupling in a proper charge transport theory framework, advancing this understanding further.

Understanding the interface effects on charge transport offers insight into advanced dielectrics design and implementation. Firstly, grasping the synergy of mechanisms underlying interface effects helps in devising new targets in material modification. For example, abundant cellulose-oil interfaces and a several-fold increase in dielectric strength of cellulose dielectrics can be realized by simple impregnation. Similarly, outstanding dielectric capacitive performance can also be acheived from nanoscale interfaces introduced by deposition layers[28].

Additionally, the application of machine learning shows immense potential for designing dielectric materials[30], which can be further unleashed greatly by the identifying the dominant factors of charge transport. Secondly, the multiscale insights help with practical modification implementation and developing valuable strategies. Recognizing the importance of electronic coupling expands modification methods and offers more flexible approaches. Its dominant role in interface effect is particularly applicable for interfaces constructing in all-organic composites, for which a single factor of charge attracting ability can be limited due to the similar all-organic components[1,6,34], as reflected by the mechanism of interface 2 in this work. Besides, the dual-interface effect on charge trapping regulation is inspirational for both two mainstream interface design categories, where the two interacted interfaces can be correspondingly realized by surface modification of nanoparticles[12,19-23] and tailoring multi-layer structures[12,24-28]. Thirdly, from a methodological perspective, this work presents an effective approach to studying charge transport in complex molecular systems, offering thorough references including theory, methods, practice, and details. Integrating multiple methods enhances flexibility and insight, making this multiscale charge transport simulation a promising approach for exploring advanced energy-related dielectrics.

## 4. Methods

### 4.1. Interfacial morphology construction by molecular dynamics

The cellulose-oil interfacial model consists of oil and repeating microfibrils. The molecular dynamics (MD) is carried out with GROMACS package[63] and OPLS-AA force field[64], velocity-rescale thermostat (time constant 0.2 ps) and Berendsen barostat (time constant 0.5 ps). The MD construction process of microfibril is shown in Fig. 1b. Firstly, the crystal cellulose nanofibril core of the microfibrils is constructed by the Cellulose Builder package[60]. Next, the crystal core is combined with the surrounding hemicellulose to form microfibril. The detailed MD relaxation process of microfibril refers to references 54 to 57. Firstly, the heterogeneous structure is first energy-minimized. This is followed by constant volume and temperature (NVT) ensemble for 5 ns with thermostat set to 450K. Next, without changing the thermostat, the barostat is set to 10 bar (NPT), and the atoms are simulated for 10 ns. Finally, the structure is relaxed at 300 K under 1 bar for another 10 ns. The time step is 0.002 ps. An increased temperature typically accelerates the simulation and in this case helps disordered hemicellulose to find its equilibrium position. However, to not disordered the crystalline cellulose, its atoms are position-restrained when the system is at high temperature. Then, the oil molecules are combined with the repeating microfibrils, and the disordered oil region is relaxed referring to references [61] and [62] while the other part is restrained. The relaxation process involves 5 ns under NVT ensemble of 450K, 10 ns under NPT ensemble of 450K and 10bar, and 5 ns under NPT ensemble of 300K and 1bar. Finally the whole interfacial system is relaxed at 300 K under 1 bar for 10 ns.

### 4.2. Related quantum chemistry calculations

For all the molecules/segments used in this work, DFT calculations (geometry optimization and single-point calculations) are performed with the hybrid PBE0[65] functional with D3BJ dispersion correction and the def2-tzvp[66] basis set using the ORCA quantum chemical package[67]. Additionally, the analysis and visualization of results are based on the software Multiwfn[68].

### 4.3. Marcus rate calculation and multiscale integration

After the interfacial morphology is constructed, it is then partitioned into hopping sites and the Marcus charge hopping rates between sites pairs are calculated separately based on equation (1), where multiscale factors are integrated. Then the charge transport dynamics can be realized by the kinetic Monte Carlo method. This stepped multiscale process is mainly carried based on the open-source software VOTCA-XTP[49]. This software provides interfaces for mainstream molecular dynamics and quantum chemistry calculation packages, realizing a complete solution for charge transport simulation in complex molecular systems.

Based on Marcus theory[38,39], the charge hopping rate $\omega_{ij}$ from site $i$ to $j$ can be calculated according to the equation (1), where $J_{ij}$ is the electronic coupling between the initial and final electronic states, $\lambda_{ij}$ is the reorganization energy, $\Delta E_{ij}$ is the site-energy difference, $\hbar$ is the reduced Planck's constant, $k_b$ is the Boltzmann constant and $T$ is the temperature. The formula (1) captures and combines multiscale influence factors by $J_{ij}$, $\lambda_{ij}$ and $\Delta E_{ij}$, which are all obtained from first-principles calculations in this study.

The site energy difference, $\Delta E_{ij}=E_i-E_j$, drives the charge transfer between site $i$ and $j$. Multiscale factors are introduced and combined together in this work through summation of all contributions due to internal energy differences, externally applied electric field, electrostatic interactions, and polarization effects, as shown in equation (3):

$$\begin{aligned}\Delta E_{ij} &= \Delta E_{ij}^{int} + \Delta E_{ij}^{env} + \Delta E_{ij}^{ext} \\ &= (\Delta E_{ij}^{DFT}) + (\Delta E_{ij}^{el} + \Delta E_{ij}^{pol}) + (\Delta E_{ij}^{ext})\end{aligned} \quad (3)$$

Note that in section 2.3, we have identified the site energy difference with the difference of ionization potentials in case of hole transport. In case of electron transport, this would correspond to the difference of the negatives of the electron affinities.

The internal energy difference $\Delta E_{ij}^{int}$ is the contribution at the quantum scale and can be calculated using DFT according to

$$\Delta E_{ij}^{int} = \Delta E_{ij}^{DFT} = E_i^{int} - E_j^{int} = \Delta U_i - \Delta U_j = (U_i^{cC} - U_i^{nN}) - (U_j^{cC} - U_j^{nN}), \quad (4)$$

where $U_i^{cC(nN)}$ is the total energy of site (segment) $i$ in the charged (neutral) state and geometry obtained from DFT. In similar fashion the reorganization energy is calculated based on Nelsen's four-point method[69]:

$$\lambda_{ij} = \lambda_i^{cn} + \lambda_j^{nc} = U_i^{nC} - U_i^{nN} + U_j^{cN} - U_j^{cC} \quad (5)$$

The effects of the environment on the energy differences, $\Delta E_{ij}^{env}$, is in our microelectrostatics approach composed of static ($\Delta E_{ij}^{el}$) and polarization ($\Delta E_{ij}^{pol}$) contributions. The electrostatic contribution to the energy of site *i* is determined from atomic partial charges as

$$E_i^{el} = \frac{1}{4\pi\varepsilon_0} \sum_{a_i} \sum_{\substack{b_k \\ k \neq i}} \frac{(q_{a_i}^c - q_{a_i}^n) q_{b_k}^n}{r_{a_i b_k}}, \quad (6)$$

where $r_{a_i b_k}$ is the distance between atoms $a_i$ and $b_k$ and the $q_{a_i}^{n(c)}$ are partial charges of atom *a* in segment *i* in state *n* or *c*. Polarization effects are incorporated at atomistic resolution. Specifically, here we further use a polarizable force field which atomic dipole polarizabilities. The induced dipoles $\mu_{a_i}^{(k)}$ are obtained iteratively by

$$\boldsymbol{\mu}_{a_i}^{(k+1)} = \omega \boldsymbol{F}_{a_i}^{(k)} \alpha_{a_i} + (1-\omega) \boldsymbol{\mu}_{a_i}^{(k)}, \quad (7)$$

where $F_{a_i}^{(k)}$ is the evaluated electric field at atom *a* in molecule/segment *i* by all atomic partial charges and induced moments and $\alpha_{a_i}$ is the isotropic atomic polarizability. The parameter ω is a damping constant for successive over-relaxation. All details about the related calculations can be found in reference 49, including the detailed expression of the $E_j^{pol}$. Finally, $\Delta E_{ij}^{ext}$ is the contribution due to the external electric field **F**

$$\Delta E_{ij}^{ext} = q(\boldsymbol{F} \cdot \boldsymbol{r}_{ij}) \quad (8)$$

*q* is the charge and $r_{ij}$ is a vector connecting site *i* and *j*. Long-range electrostatic interactions are accounted for via a periodic embedding of aperiodic excitations based on Ewald summation[50]. Polarization effects are considered within a cutoff of 3.0 nm around each individual segment.

The electronic coupling $J_{ij}$ expresses the coupling strength between two electronic states localized on sites *i* and *j*, respectively, and is defined as

$$J_{ij} = \langle \varphi_i | \hat{H} | \varphi_j \rangle, \quad (8)$$

where the $\varphi_i$ and $\varphi_j$ are the molecular orbital wave functions of the related electronic states, respectively, and the $\hat{H}$ is the Hamiltonian of the dimer. Within the frozen-core approximation, the usual choice for the diabatic wave functions are the frontier orbitals. Equation (8) is evaluated in this work using the dimer projection method[48].

21. Zhang, X., Li, B. W., Dong, L., Liu, H., Chen, W., Shen, Y., & Nan, C. W. (2018). Superior energy storage performances of polymer nanocomposites via modification of filler/polymer interfaces. Advanced Materials Interfaces, 5(11), 1800096.

22. Nilagiri Balasubramanian, K. B., & Ramesh, T. (2018). Role, effect, and influences of micro and nano-fillers on various properties of polymer matrix composites for microelectronics: a review. Polymers for Advanced Technologies, 29(6), 1568-1585.

23. Shen, Z. H., Wang, J. J., Lin, Y., Nan, C. W., Chen, L. Q., & Shen, Y. (2018). High-throughput phase-field design of high-energy-density polymer nanocomposites. Advanced Materials, 30(2), 1704380.

24. Wang, Y., Cui, J., Yuan, Q., Niu, Y., Bai, Y., & Wang, H. (2015). Significantly enhanced breakdown strength and energy density in sandwich-structured barium titanate/poly (vinylidene fluoride) nanocomposites. Advanced Materials, 27(42), 6658-6663.

25. Baer, E., & Zhu, L. (2017). 50th anniversary perspective: dielectric phenomena in polymers and multilayered dielectric films. Macromolecules, 50(6), 2239-2256.

26. Qiao, Y., Yin, X., Zhu, T., Li, H., & Tang, C. (2018). Dielectric polymers with novel chemistry, compositions and architectures. Progress in Polymer Science, 80, 153-162.

27. Feng, M., Feng, Y., Zhang, T., Li, J., Chen, Q., Chi, Q., & Lei, Q. (2021). Recent advances in multilayer-structure dielectrics for energy storage application. Advanced Science, 8(23), 2102221.

28. Cheng, S., Zhou, Y., Li, Y., Yuan, C., Yang, M., Fu, J., He, J., & Li, Q. (2021). Polymer dielectrics sandwiched by medium-dielectric-constant nanoscale deposition layers for high-temperature capacitive energy storage. Energy Storage Materials, 42, 445-453.

29. Liu, G., Lei, Q., Feng, Y., Zhang, C., Zhang, T., Chen, Q., & Chi, Q. (2023). High-temperature energy storage dielectric with inhibition of carrier injection/migration based on band structure regulation. InfoMat, 5(2), e12368.

30. Meng, Z., Zhang, T., Zhang, C., Shang, Y., Lei, Q., & Chi, Q. (2023). Advances in Polymer Dielectrics with High Energy Storage Performance by Designing Electric Charge Trap Structures. Advanced Materials, 2310272.

31. Li, S., Xie, D., & Lei, Q. (2020). Understanding insulation failure of nanodielectrics: tailoring carrier energy. High Voltage, 5(6), 643-649.

32. Tan, D. Q. The search for enhanced dielectric strength of polymer-based dielectrics: a focused review on polymer nanocomposites. Journal of Applied Polymer Science. 2020, 137(33), 49379.

## Acknowledgments

This work was supported by the National Natural Science Foundation of China, General Project, No. 52477027.

## Author contributions

H.X.Z. conceived the idea. H.X.Z. designed the simulations. H.X.Z. performed the simulations. H.X.Z. and B.B. analyzed the data. H.X.Z., L.X.A. and B.B. wrote the manuscript. All authors discussed the results and commented on the manuscript.

## Competing interests

The authors declare no competing interests.

# Supplementary Information

Synergistic Interface Effects in Composite Dielectrics: Insights into Charge Trapping Regulation through Multiscale Modeling

*Haoxiang Zhao[1,2,3], Lixuan An[4], Daning Zhang[1], Xiong Yang[1], Huanmin Yao[1], Guanjun Zhang[1], Haibao Mu[1]\* and Björn Baumeier[2,5]\**

1 State Key Laboratory of Electrical Insulation and Power Equipment, School of Electrical Engineering, Xi'an Jiaotong University, Xi'an 710049, China

2 Department of Mathematics and Computer Science, Eindhoven University of Technology, Eindhoven 5600MB, The Netherlands

3 Department of Applied Physics, Eindhoven University of Technology, Eindhoven 5600MB, The Netherlands

4 KERMIT, Department of Data Analysis and Mathematical Modelling, Ghent University, Ghent 9000, Belgium

5 Institute for Complex Molecular Systems, Eindhoven University of Technology, Eindhoven 5600MB, The Netherlands

\*Corresponding author

E-mail address:

haibaomu@mail.xjtu.edu.cn (Haibao Mu)

b.baumeier@tue.nl (Björn Baumeier)

## 1. Molecule structures of cellulose and hemicellulose.

The cellobiose, a reducingsugar, is a disaccharide with the formula C12H22O11 and consists of two D-glucose molecules linked by a $\beta(1\rightarrow4)$ bond as shown in Fig. S1a. The cellulose molecular chains are made up of 20 linked D-glucose in this way as shown in Fig. S1b. The detailed structure of the repeating unit of hemicellulose is shown in Fig. S1c. This branched polysaccharide is composed of D-glucose, D-galactose, and D-mannose residues. As the linking core, the mannose is $\beta(1\rightarrow4)$ linked to a D-glucopyranose residue, while simultaneously forming an $\alpha(1\rightarrow6)$ glycosidic bond with a D-galactopyranose unit, creating a branch at the C6 position of the mannose. In this way, the hemicellulose is formed by 10 repeating trisaccharide units as shown in Fig. S1d. The choose of the molecule length is following the previous related work, where the molecule length are studied with related structure conformation and physicochemical properties[1-4].

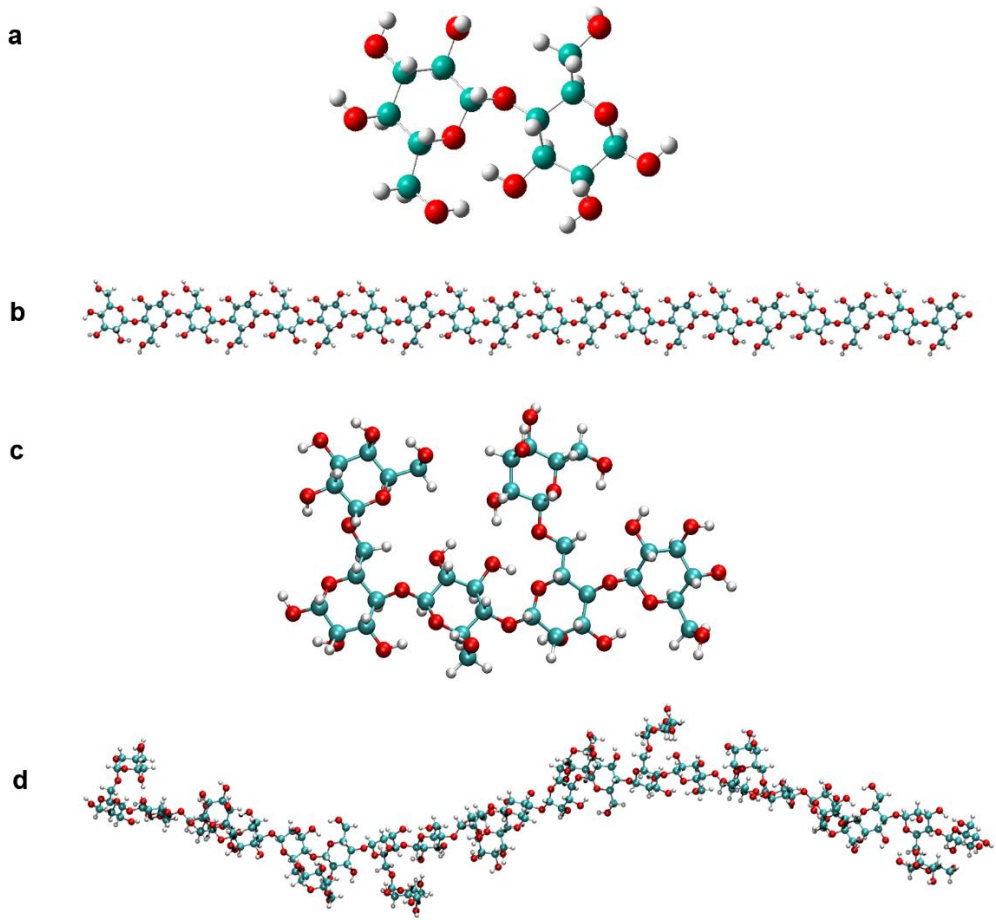

**Figure S1.** Molecule structure of cellulose and hemicellulose. **a** Cellobiose. **b** Cellulose molecule. **c** Repeating unit of hemicellulose. **d** hemicellulose molecule.

## 2. Results of contribution of $\Delta E_{ij}$ to $\omega_{ij}$, contribution of $J_{ij}$ to $\omega_{ij}$, and the correlations.

Based on the description in section 2.4, the Marcus rate in equation (1) is split into two parts: the exponential term $G_{ij}(\Delta E_{ij})$ and the remaining factor $F_{ij}(J_{ij})$, as shown in equation (2) and (3) below. $G_{ij}(\Delta E_{ij})$ and $F_{ij}(J_{ij})$ represent the contributions of site energy difference $\Delta E_{ij}$ and electronic coupling $J_{ij}$ to $\omega_{ij}$, respectively. The statistical region-based results of $G_{ij}(\Delta E_{ij})$ and $F_{ij}(J_{ij})$, and the correlation analysis to the hopping rates shown in Fig. S2 to Fig. S5.

$$\omega_{ij} = \frac{2\pi}{\hbar} \frac{J_{ij}^2}{\sqrt{4\pi\lambda_{ij}k_B T}} \exp\left[-\frac{(\Delta E_{ij} - \lambda_{ij})^2}{(4\lambda_{ij}k_B T)}\right] = F_{ij}(J_{ij}) \cdot G_{ij}(\Delta E_{ij}) \qquad (1)$$

$$F_{ij}(J_{ij}) = \frac{2\pi}{\hbar} J_{ij}^2 \tag{2}$$

$$G_{ij}(\Delta E_{ij}) = \frac{\exp\left[-\frac{(\Delta E_{ij} - \lambda_{ij})^2}{(4\lambda_{ij}k_{\mathrm{B}}T)}\right]}{\sqrt{4\pi\lambda_{ij}k_{\mathrm{B}}T}} \tag{3}$$

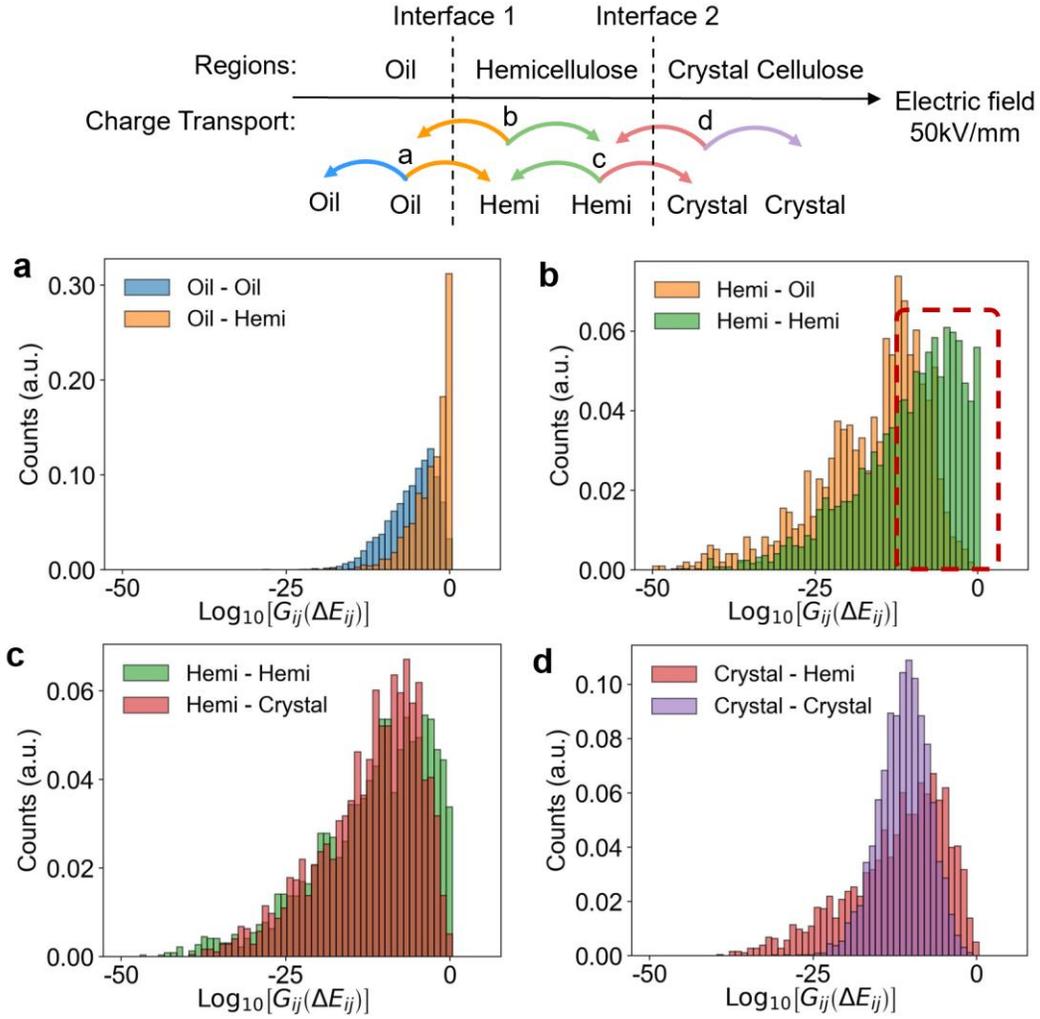

**Figure S2.** Statistical region-based results of $G_{ij}(\Delta E_{ij})$, the $\Delta E$ contribution to hopping rate $\omega_{ij}$

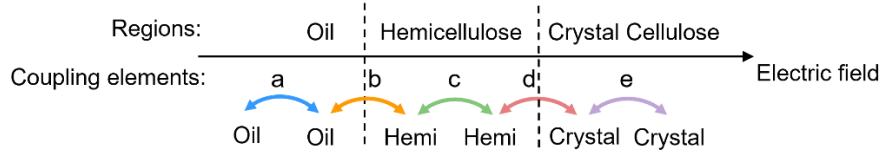
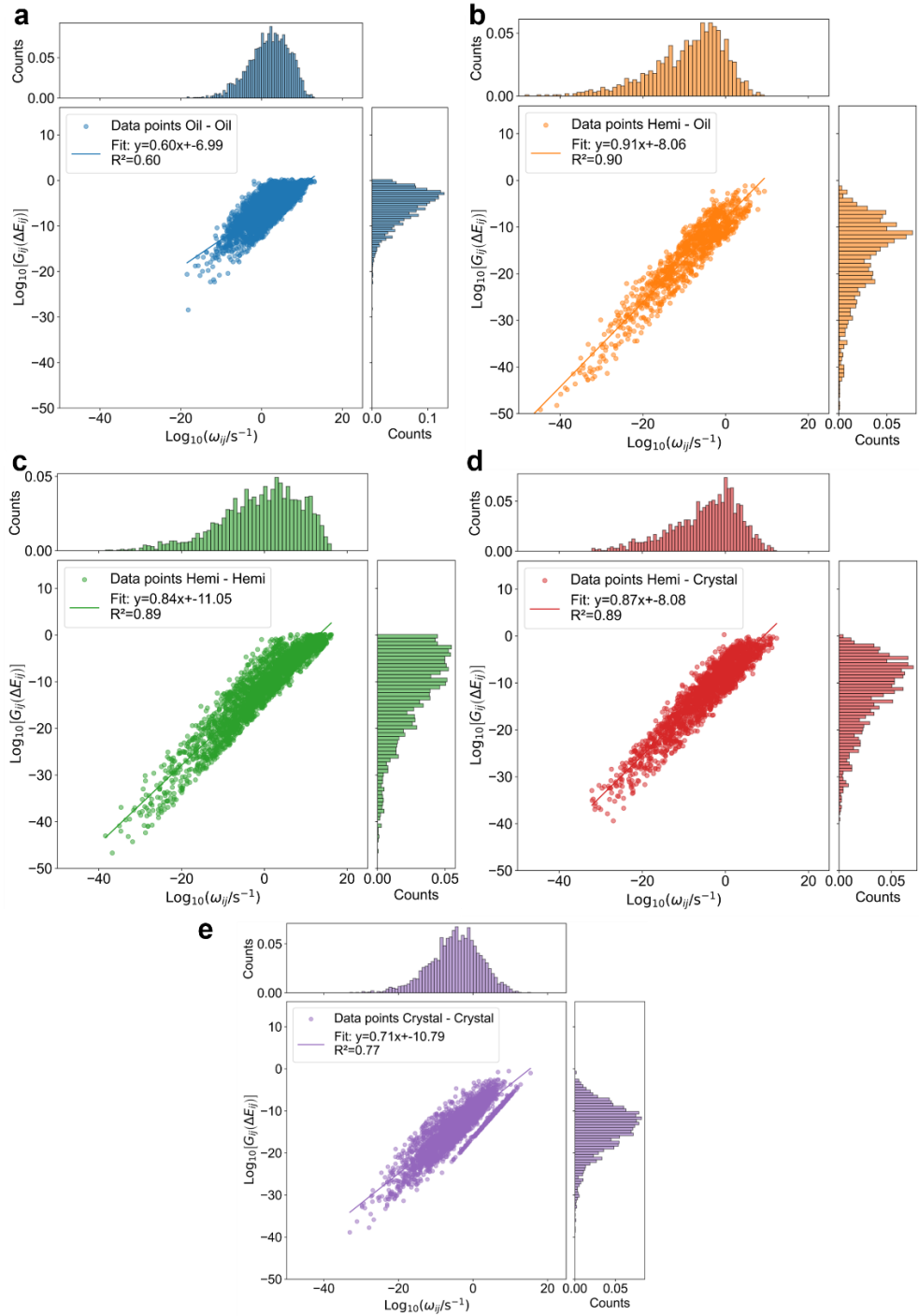

**Figure S3.** Correlation between the $\Delta E$ contribution $G_{ij}(\Delta E_{ij})$ and hopping rate $\omega_{ij}$

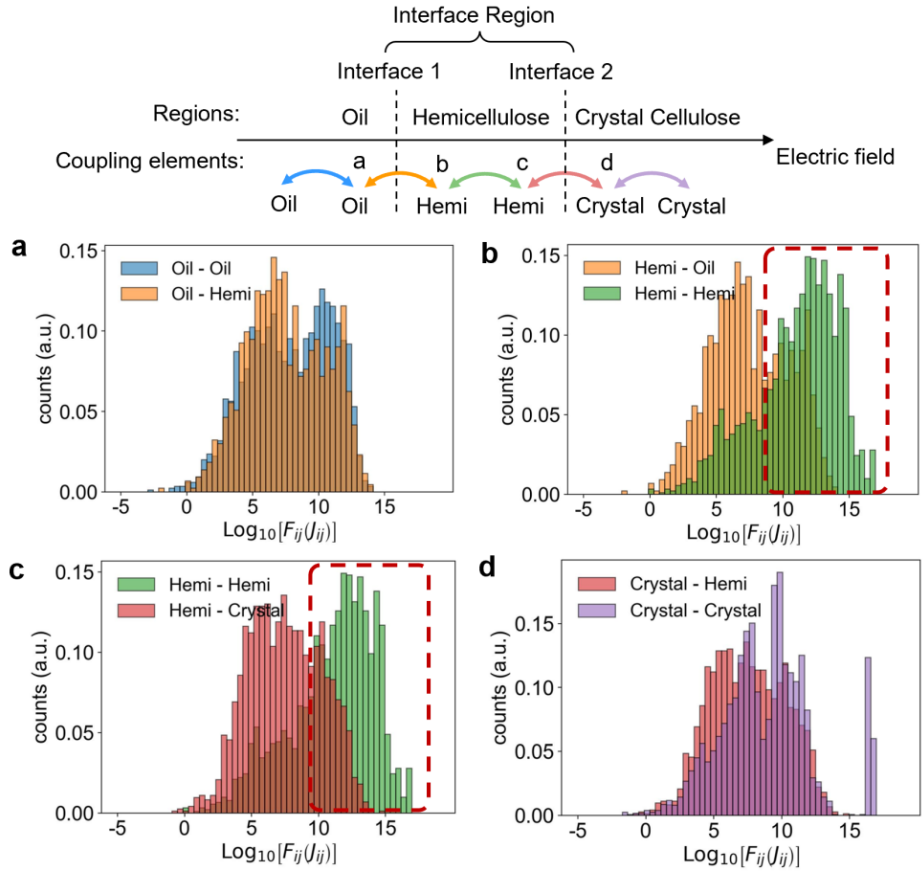

**Figure S4.** Statistical region-based results of $F_{ij}(J_{ij})$, the $J_{ij}$ contribution to hopping rate $\omega_{ij}$

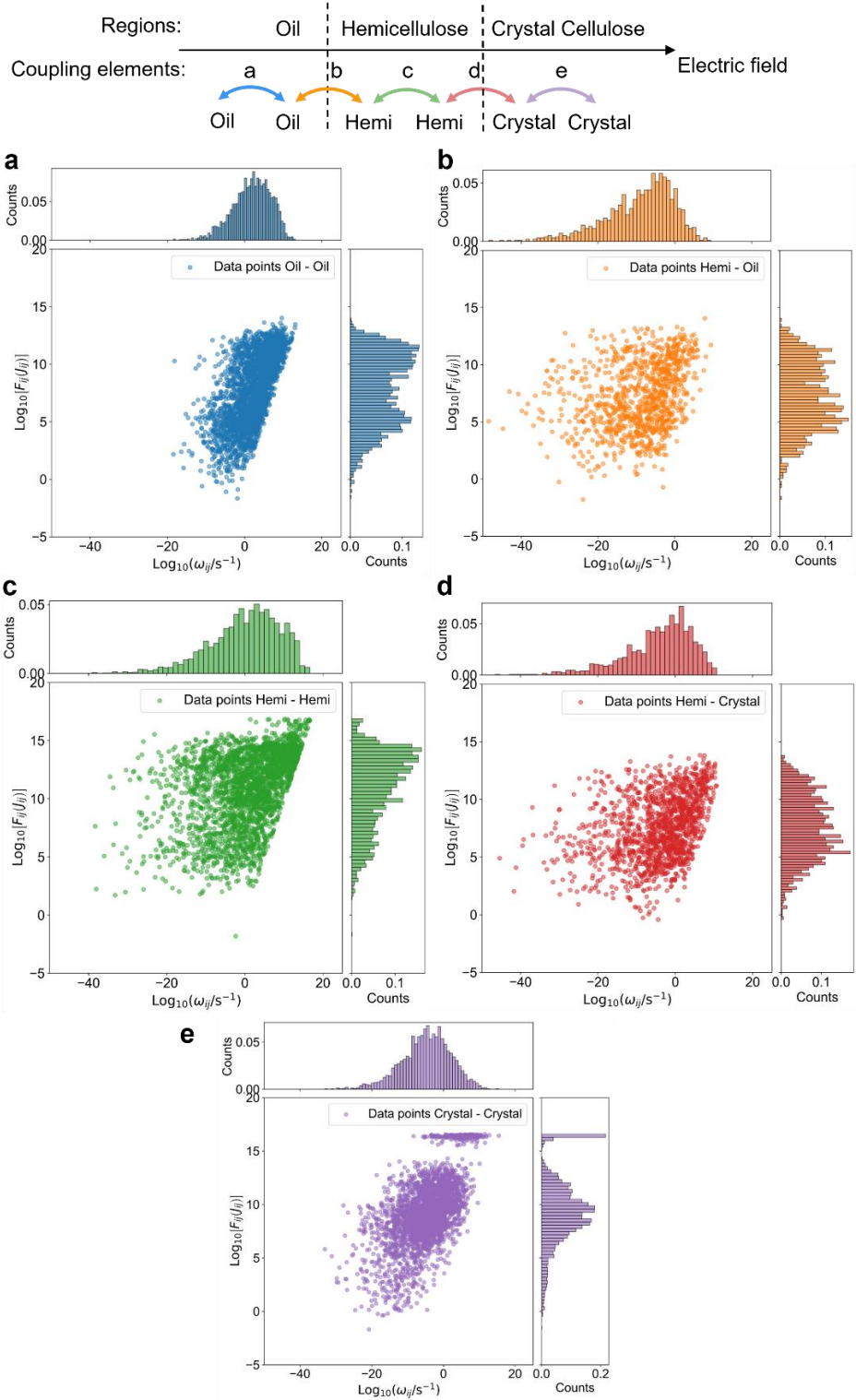

**Figure S5.** Correlation between $J_{ij}$ contribution $F_{ij}(J_{ij})$ and hopping rate $\omega_{ij}$. There is no obvious correlation comparing to the results in Fig. S3.